\documentclass[12pt]{iopart}
\usepackage{color}
\usepackage{graphicx}

\usepackage{iopams}

\begin{document}
\title{Conductance microscopy of quantum dots weakly or strongly coupled to the conducting channel}

\author{K. Kolasi\'nski and B. Szafran}
\address{AGH University of Science and Technology, Faculty of Physics and Applied Computer Science,\\
al. Mickiewicza 30, 30-059 Krak\'ow, Poland}

\begin{abstract}
We consider scanning gate conductance microscopy of an open quantum dot that is connected to the conducting channel
using the wave function description of the quantum transport and a finite difference approach.
We discuss the information contained in conductance ($G$) maps.
We demonstrate that the maps for a delta-like potential perturbation
exactly reproduce the local density of states for the quantum dot that is weakly coupled to the channel,
i.e. when the connection of the channel to the dot transmits
a single transport mode only. We explain this finding in terms
of the Lippmann-Schwinger perturbation theory. We demonstrate that the signature of the weak coupling conditions is the conductance
which for $P$ subbands at the Fermi level varies between $P-1$ and $P$ in units of  ${2e^2}/{h}$.
For stronger coupling of the quantum dot to the channel the $G$ maps resolve the local density of states
only for very specific work points with the Fermi energy coinciding with quasi-bound energy levels.
\end{abstract}


\maketitle
\section{Introduction}

Scanning gate microscopy (SGM) \cite{review1,review2} is an experimental technique that
probes the reaction of confined electron system to perturbation introduced by the charge of atomic force microscope
tip that sweeps above the sample area. The SGM technique has been applied to a number of
systems, including quantum point contacts \cite{qpc0,qpc1,qpc2,qpc3,qpc4,qpc5,qpc6,qpc7,qpc8}, currents  in the quantum Hall regime \cite{qh1,qh2,edge},
quantum rings \cite{review1,pala1,pala2,pala3,szafran,chwiej,petrovic}, and quantum dots \cite{review2,qdc1,shulz1,shulz2,qda1,g6}.
The quantum dots that are studied with SGM include the closed ones \cite{qdc1}, in which the transport is governed
by the Coulomb blockade of the current. The potential of the tip changes the energy of the
confined system in an extent that depends on the local charge density. The energy shift can be experimentally
measured which allows for the read-out of the confined charge density distribution \cite{qdc1}.
On the other hand, open quantum dots \cite{review2} are strongly coupled to the reservoirs, the flow of the current is never strictly blocked,
and the tip potential only modifies the conductance. According to the Landauer-B\"uttiker theory the coherent conductance is determined by the properties of Fermi level wave functions \cite{datta}.
In particular, the conductance maps of systems based on quantum point contacts are generally well
correlated to the current flow at the Fermi level \cite{qpc0,qpc2,qpc4,qpc5,qpc6,qpc7,kramer,bruno,cresti}.
For open quantum dots and quantum rings the SGM signal was
attributed \cite{review1,pala1,pala2,pala3,shulz1} to the electron density at the Fermi level called the local density of states (LDOS).
According to the semi-classical WKB approximation the perturbation -- due to the tip for instance --
changes locally the electron wavelength. The wavelength modulation should have the largest effect in the areas of large LDOS and thus the latter should be resolved in conductance maps.

Arguments against interpretation of conductance maps as images of the current flow or LDOS in the absence
of the tip were risen along with formulation of the Lippmann-Schwinger theory of SGM \cite{pert1,pert2}. In terms of the
perturbation theory \cite{pert1,pert2} the corrections to conductance due to the tip are non-local as expressed
by the wave function which is a global field.
Interference effects introduced by the tip are observed in SGM experiments
with quantum point contacts \cite{qpc1,qpc4,abbout,jura,kozikov}.

In this work we discuss the interpretation of the $G$ mapping for the quantum dot side-attached to a conducting channel.
We develop the results of our previous study \cite{kolasinski} for a cavity strongly
coupled to the channel and a single mode transport which indicated that the LDOS and $G$ maps are clearly correlated only at Fano resonances \cite{shulz3,morfonios,clerk,Barn,Miro}.
Here, we study a quantum dot with variable opening (coupling) to the channel.
We focus on the limit of a weak point-like perturbation. In this way we determine the
best possible spatial resolution of conductance maps and its possibly maximal correlation with the unperturbed local density of states.
In quantum dots that are strongly coupled to the channel, the SGM map is usually far from LDOS even for the weak point-like perturbation.
Nevertheless, we find that for weakly coupled quantum dots the SGM conductance maps
are highly correlated to the LDOS for any Fermi energy. We explain this finding and indicate
the experimental conditions in which SGM measurements precisely probe the local density of states.

\section {Model}
\begin{figure}[!htb]
\includegraphics[scale=.3]{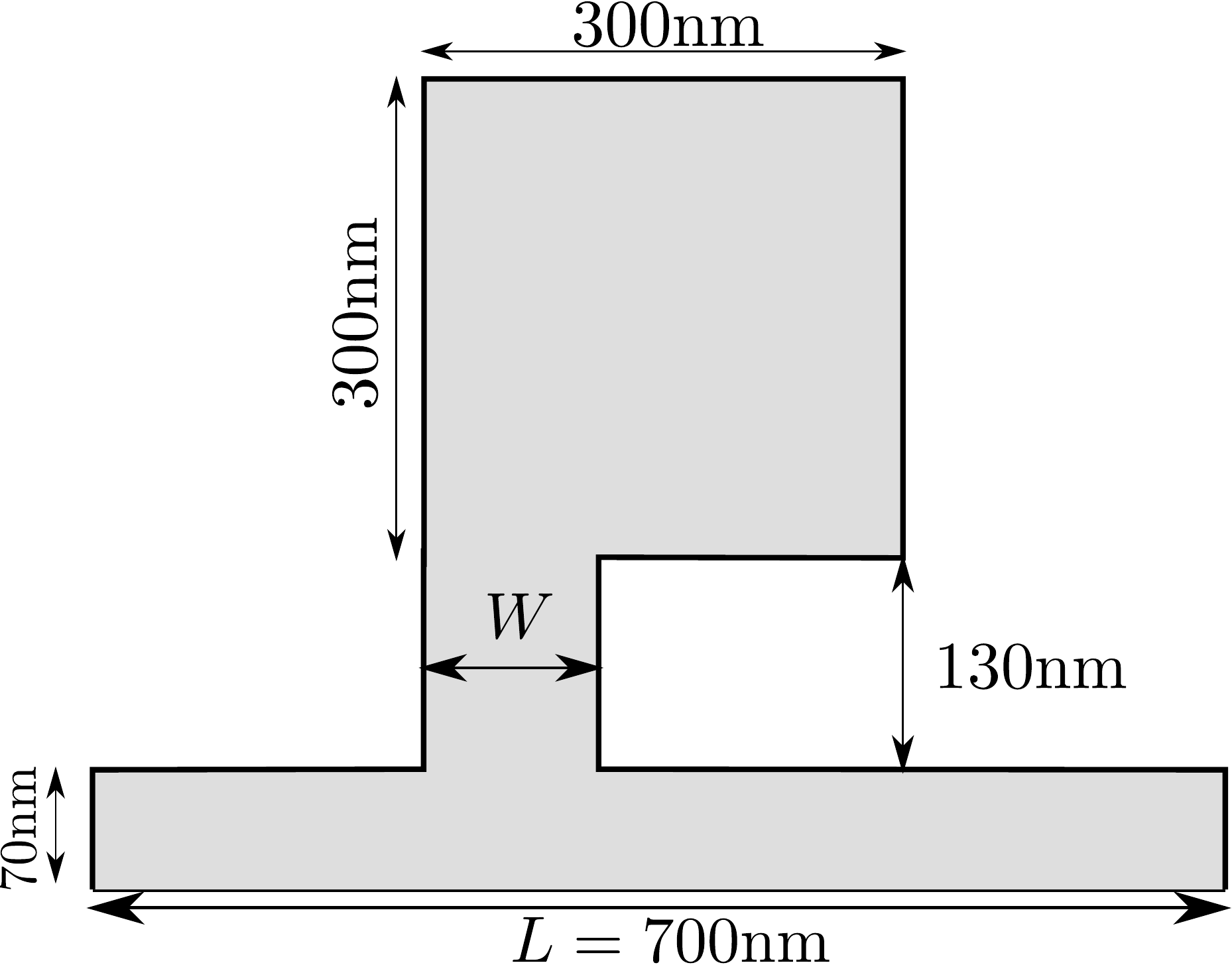}
\caption{Schematics of the considered system with 70 nm wide channels with a quantum dot
of dimensions 300 nm $\times$ 300 nm that opens to the channel by a 130 nm long connection of width $W$.
The computational box contains $L=700$ nm segment of the channel. \label{Wneka}}
\end{figure}

\begin{figure}[!htb]
\begin{centering}
\includegraphics[scale=.4]{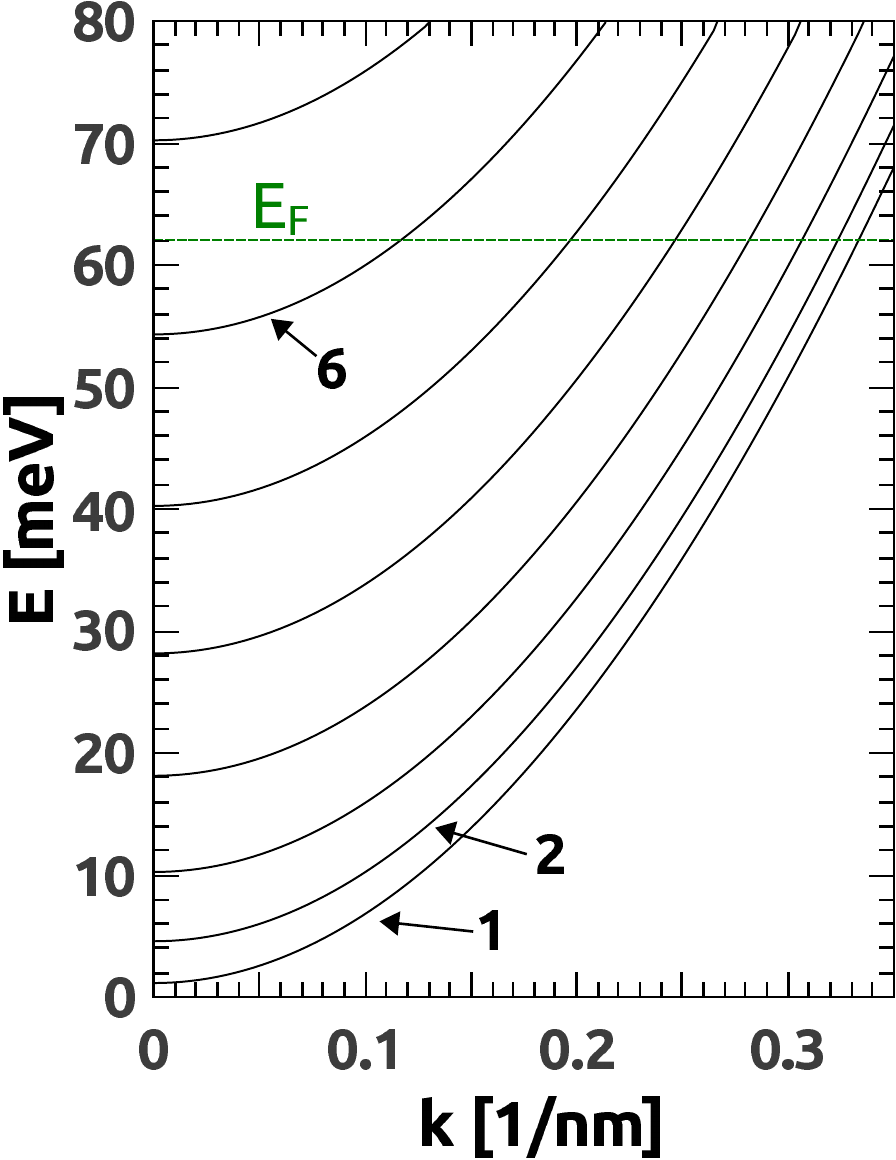} %
\end{centering}
\caption{Dispersion relation for the channel that is 70 nm wide. The numbers of subbands are indicated.
The horizontal green line shows the  Fermi energy level
for $P=6$ subbands participating in the current flow.}\label{dr}
\end{figure}

We consider the system that is depicted in Fig. \ref{Wneka} with a square $300$ nm $\times$ 300 nm quantum dot
that opens to the 70 nm wide channel by a connection that is 130 nm long and has a width of $W$.
The width of the opening $W$ that determines the strength of the coupling of the quantum dot to the channel
 is the central parameter discussed in this work. For $W<300$ nm the dot is asymmetrically attached
 to the channel.

We apply a strictly two-dimensional model of the system and use the effective
mass Hamiltonian
\begin{equation} H=-\frac{\hbar^2}{2m}\nabla^2+V_c(x,y)+V_t(x,y;x_t,y_t),\end{equation}
where $m=0.067m_0$ is GaAs electron effective mass, $V_c$ is the confinement potential and $V_t$ is the potential of the tip.
For $V_c$ we use a quantum well potential with $V_c=0$ inside the quantum dot and the channel (gray area in Fig. 1)
and $V_c\rightarrow \infty$ in the rest of the plane.

In the channel far away from the scattering region the Hamiltonian eigenstates can be put in a separable form $\Psi_n^k(x,y)=\exp(ikx) \psi_n(y)$,
with the wave vector $k$ and the subband quantum number $n$.
The dispersion relation for the channel is plotted in Fig. \ref{dr}. In this work we consider up to $P=6$ subbands participating
in the current flow at the Fermi level.
The conductance is evaluated using the Landauer formula
\begin{equation} G=\frac{2e^2}{h} T=\frac{2e^2}{h} \sum_{p=1}^P\sum_{q=1}^P T_{pq},\end{equation}
where the factor of 2 accounts for the spin-degeneracy, and $T_{pq}$ is the electron transfer probability from $p$-th subband of the input
lead to $q$-th subband of the output lead. The transfer probabilities are extracted from the Schroedinger equation that is solved using the finite
difference method with the gauge-invariant kinetic energy discretization \cite{governale}.
The computational box covers a section of the channel of length $L=700$ nm  which is large enough to neglect the evanescent modes appearing near the scattering
region at the ends of the channel.
The boundary conditions at these ends are then set by superposition of current carrying modes.
For the electron incident from the left lead in subband $p$ at the input channel
one finds the reflected electron waves in all the subbands \begin{equation}\Psi_{in}(x,y)= c_p \exp(ik_px)\psi^{k_p}_p(y)+\sum_{q=1}^P d_q \exp(-ik_qx)\psi^{-k_q}_q(y),\end{equation} where $c_p$ and $d_q$ are the amplitudes
of the incident and reflected waves. At the right end of the box we find the wave function scattered to all the subbands \begin{equation} \Psi_{out}(x,y)= \sum_{q=1}^P e_q \exp(ik_qx)\psi^{k_q}_q(y).\end{equation}
The scattering amplitudes $c_p,d_p,e_p$ are calculated using a self-consistent scheme of Ref. \cite{szafran}.
When the convergence of the scheme is reached \cite{szafran}, the transfer probability can be evaluated as \begin{equation} T_{pq}=\left|\frac{e_q}{c_p}\right|^2 \frac{k_q}{k_p}. \end{equation}

The effective tip potential that results from
the screening of the tip charge  by the deformation of the two-dimensional electron gas
was evaluated by Schroedinger-Poisson calculations \cite{szafran,chwiej,kolasinski}.
 The effective tip potential
turns out to be close to a Lorentzian \cite{szafran}. Here, we consider the point-like tip potential (see Introduction) using the Lorentz function
\begin{equation}
V_t(x,y;x_t,y_t)=\frac{Ud^2}{(x-x_t)^2+(y-y_t)^2+d^2} \label{lp}
\end{equation}
with small values of both the tip height $U=0.05$ meV and the potential width $d=4$ nm.
The small value of the latter is of a more basic importance.
In fact, all the results presented below -- with the exception of the contrast of conductance maps
remain the same when $U$ is significantly increased, by a factor of 20 for instance (see below the of Fig. \ref{p6}).
The value of $d$ in the experiments is larger, of the order of the distance between the tip and the electron gas as
we established previously using the Schr\"odinger-Poisson calculations for the tip potential screening problem \cite{szafran,chwiej,kolasinski}.
For larger $d$ the resolution of $G$ images is limited, as discussed in detail in Ref. \cite{szafran}.
In the present paper we study the limit of maximal resolution of $G$ map that is allowed by the quantum transport properties of the Fermi level electrons.

The local density of states \cite{pala2,shulz1} is a feature of the unperturbed system that we want
to extract using the scanning gate microscopy. For a given Fermi energy the local density of states is defined as the sum of the scattering wave functions. Denoting by $\Psi_p^{\pm}$  the scattering wave function
for the electron incident from subband $p$ from the left ($+$) or right ($-$) lead, the local density of states
is evaluated as \begin{equation} \mathrm{LDOS}(x,y)=\sum_{p=1}^P|\Psi_p^+(x,y)|^2+\sum_{p=1}^P|\Psi_p^-(x,y)|^2. \end{equation}
We compare  LDOS$(x,y)$ with the $G(x,y)$ maps as functions of the tip position. For comparison we normalize
both maps. For $G_{max}$ and $G_{min}$ standing for the maximal and minimal values of conductance as obtained
when the system is scanned by the tip, the normalized conductance is defined by $N(G)=(G(x,y)-G_{min})/(G_{max}-G_{min})$.
The correlation between the maps $a(x,y)$ and $b(x,y)$ is calculated as \begin{equation} r=\frac{\int_S  (a(x,y)-\langle a\rangle )(b(x,y)-\langle b\rangle )dxdy}{S\sigma_a \sigma_{b}}, \label{corel}\end{equation}
where $\langle a\rangle=\frac{1}{S}\int_S  a(x,y)dxdy$, $\sigma_a^2=\frac{1}{S}\int_S  (a(x,y)-\langle a\rangle)^2dxdy$, and $S$ is the area where the comparison is made -- here the square quantum dot area.

\section{Results}

\begin{figure*}[htbp]
\begin{centering}
\includegraphics[trim = 0 -15pt 0 0 , scale=0.5]{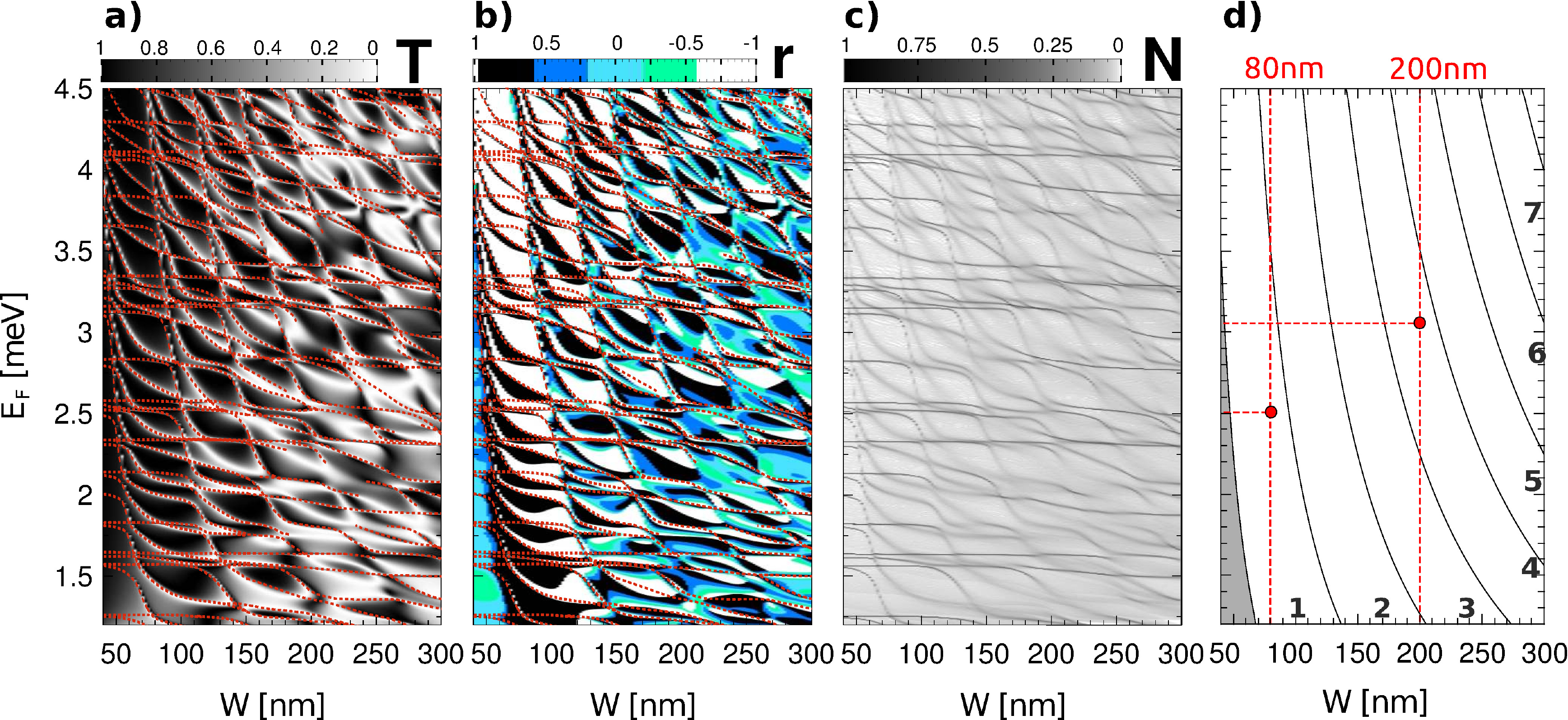}
\caption{ Single subband transport. (a) Electron transfer probability as a function of the energy and the connection width $W$ [see Fig. \ref{Wneka}]. (b) Correlation coefficient $r$ of
the local density of states [Eq. (\ref{corel})] and conductance maps for the system perturbed by the tip potential [Eq. (\ref{lp})] for $d=4$ nm and $U=0.05$ meV. (c) Density of localized states
(in arbitrary units). The dashed lines in (a) and (b) show the position of the most
pronounced resonances of panel (c).
(d) Energy levels of a 1D quantum well as function of its width $W$
($E_n=\frac{1}{2m}\left[\frac{n \hbar \pi }{W}\right]^2$).
The integers within the panel show the number of transparent modes of the connection.
Red dots indicate the work points ($E$ and $W$) which are discussed in the text.
} \label{glowny}
\end{centering}
\end{figure*}

\begin{figure*}[htbp]
\begin{centering}
\includegraphics[scale=0.5]{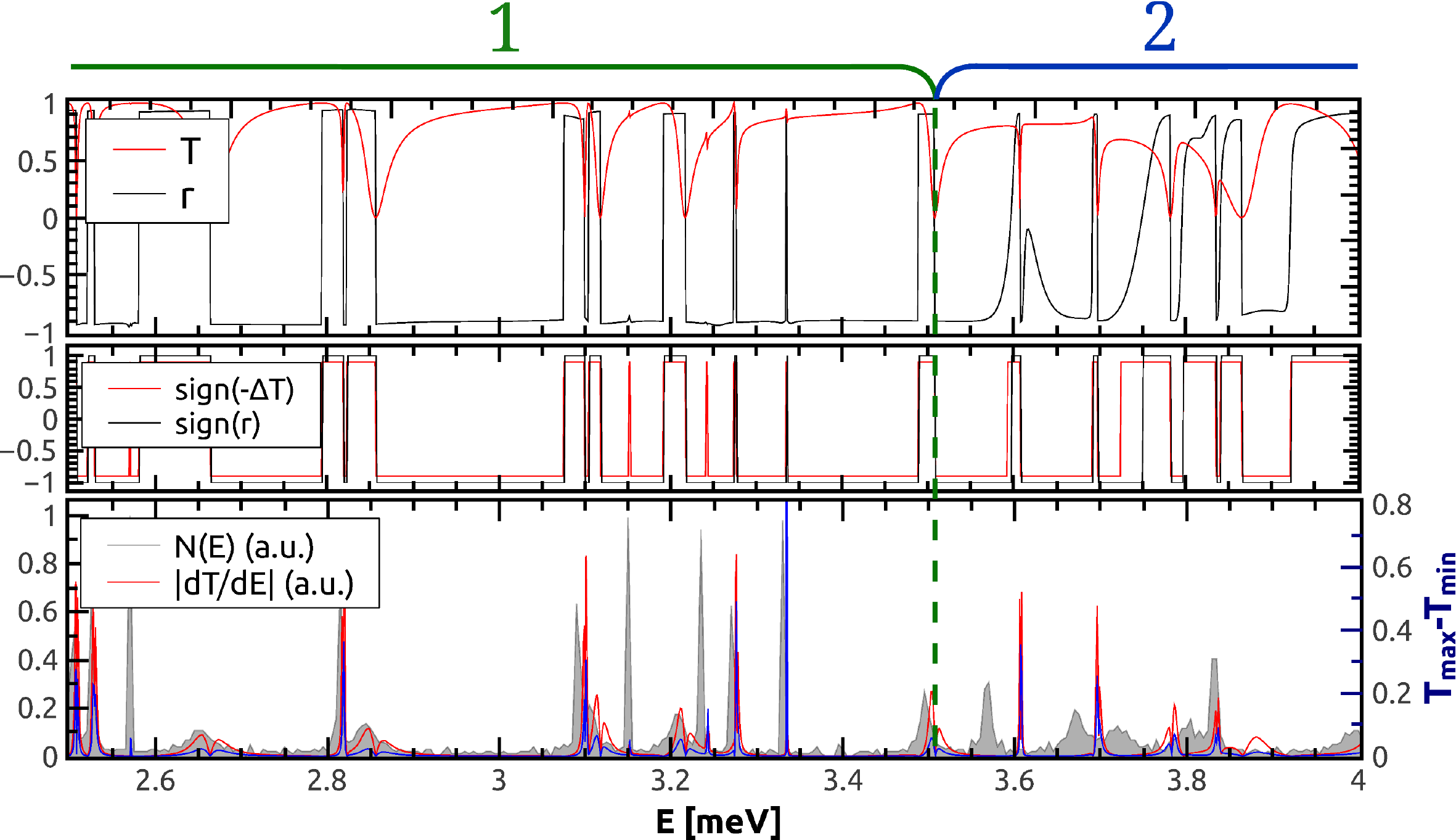}
\end{centering}
\caption{Lowest subband transport $P=1$ case. Top panel: transfer probability (red curve) -- a cross section of Fig. \ref{glowny}(a) for $W=80$ nm, $r$ -- correlation (black curve)
of the local density of states to the conductance map as obtained with the Lorentz ansatz of Eq. (\ref{lp}) -- a cross
section of Fig. \ref{glowny}(b). Middle panel: $\mathrm{sign}(r)$ (black curve) and  $-0.9\times\mathrm{sign}(dT/dE)$ (red curve). Bottom panel: the gray signal shows $N(E)$ -- the cross section of Fig. \ref{glowny}(c), the red line shows the absolute value of $dT/dE$ (left axis), and the blue line  the contrast of the $T$ map (right axis)
for a  given energy. The ranges 1 and 2 correspond to the number of subbands in the connection -- see Fig. 3(d).
} \label{jpo}
\end{figure*}

\begin{figure*}[htbp]
\begin{centering}
\includegraphics[scale=0.5]{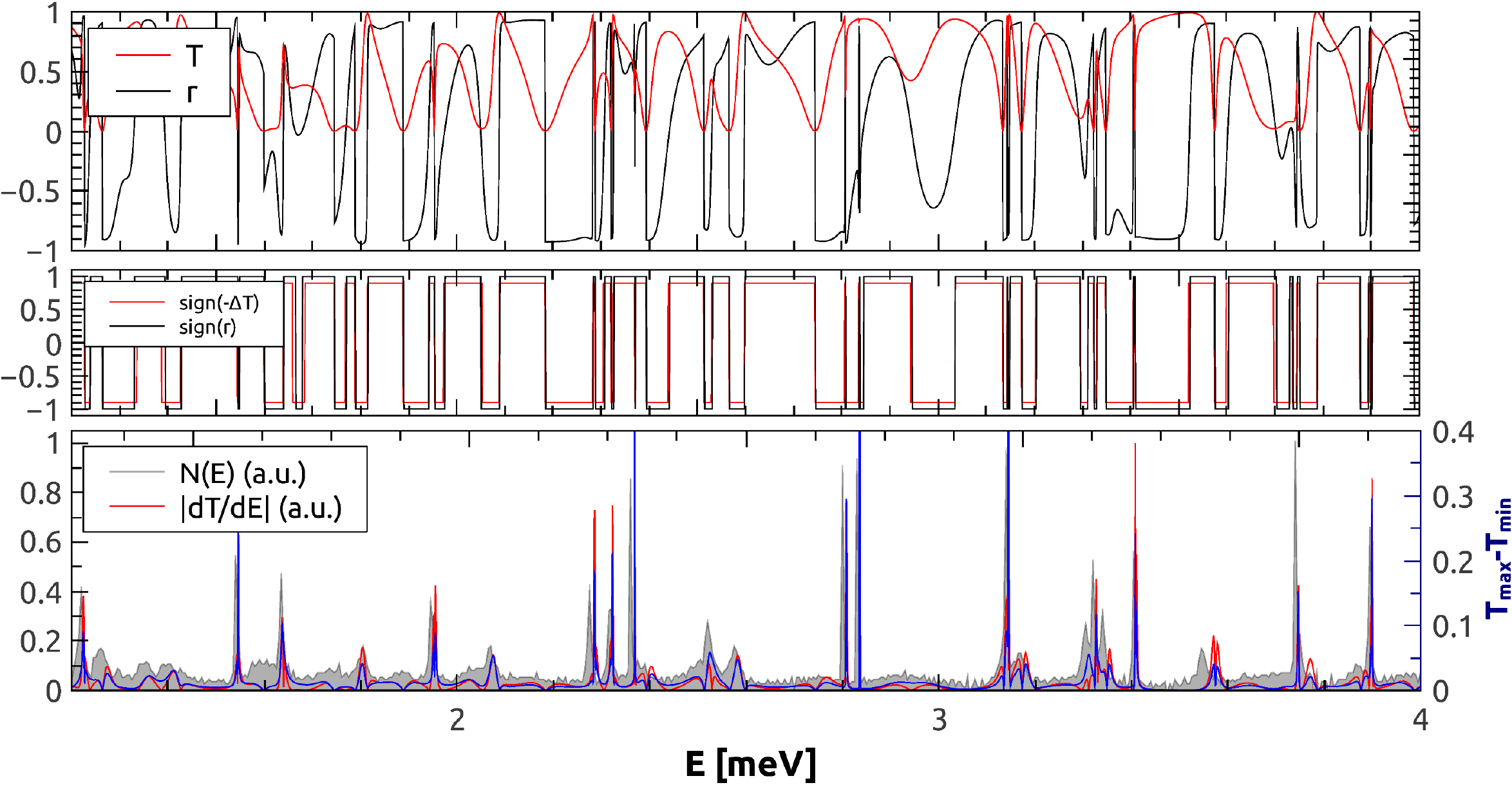}
\end{centering}
\caption{Same as Fig. \ref{jpo} only for connection width $W=200$ nm. The width of the channel is unchanged with $P=1$, but the number of transparent modes in the connection reaches 5 at the end of the horizontal
axis -- see Fig. 3(d).}\label{jpd}
\end{figure*}

\subsection{Single-subband transport}

Let us begin by characterization of the system as a scatterer for low Fermi energy with $P=1$.
Figure \ref{glowny}(a) shows the electron transfer probability as a function of the connection width $W$ for the Fermi energy range that corresponds to the transport in the lowest subband.
For small $W$ [see Fig. \ref{Wneka}] the channel is only weakly coupled to the quantum dot and $T$ is generally close to 1.
The backscattering of the incident electron involves Fano interference of the channel and quantum dot wave functions
which occurs only when the Fermi energy coincides with the energy of localized states within the dot.
We determined the density of localized states using the Mandelshtam stabilization method \cite{mandel}.
The calculated  localized states density  have been plotted in Fig. \ref{glowny}(c).
The pattern of localized energy levels has been additionally marked with the dashed lines in Figs. \ref{glowny}(a,b).
We can see that for small $W$ narrow valleys of low $T$ are formed near the quasi-bound energy levels .
For larger $W$ -- as the dot-channel coupling becomes stronger, low $T$ areas appear also off the energy positions of the localized states [Fig. \ref{glowny}(a)].

Figure \ref{glowny}(b) shows that the value of the correlation coefficient ($r$) between the LDOS and the conductance map.
We notice that generally at the resonances the correlation coefficient $r$ changes its sign.
The correlation between the $G$ map and LDOS is sensitive to the number of modes
accessible for transport in the connection between the channel and the quantum dot [see Fig. 1].
The connection of width $W$ becomes transparent for the Fermi energy which
exceeds the energy of the lateral quantization.
The number of transparent modes in the connection can be evaluated from
Fig. 3(d), where we plotted the energy spectrum of the one-dimensional infinite quantum well of width $W$,
$E_n=\frac{1}{2m}\left[\frac{n \hbar \pi }{W}\right]^2$.
The number of modes in the connection has no significant effect on the transport probability [Fig. 3(a)] but its effect
on $r$ is striking. In Fig. 3(b) we find a narrow band  of low correlation in the near the left lower corner
of the Figure (low $E$ and $W$). We can see that this region corresponds to a non-transparent connection [shaded area in Fig. 3(d)].
For zero transport modes in the connection, the wave function of the incident electron does not penetrate the quantum dot. The tip
changes the potential landscape outside the reach of the scattering wave function with no consequences for conductance.
Next, in Fig. 3(b) we can see that  $r$ takes binary values $\pm 1$ when
the connection is open for a single transport channel only [see the region marked by '1' in Fig. 3(d)].
 With opening of subsequent transport modes in the connection
 the areas of low $|r|$ become wider.
With each subsequent  channel opened in the connection the $G$ maps loose
 their correspondence to LDOS.

The top panel of Fig. \ref{jpo}  shows the cross section of Fig. \ref{glowny}(a) for $W=80$ nm,
with dips in $T$ (red line), which usually have asymmetric profiles  -- a  characteristic of the Fano interference.
The energy range marked by 1 and 2 on top of the figure correspond to the number of transparent subbands of the channel-dot connection [see Fig. 3(d)].
The dependence of $r$ on the energy is binary in the range 1. For the second range $r$ becomes a continuous function of the
energy. In this region we notice that $|r|$ tends to 1 only at the conductance resonances.
This feature is preserved also for much wider dot-channel connection supporting a large number of transport modes -- see the results for $W=200$ nm
in Fig. \ref{jpd}.

In Figs. \ref{jpo} and \ref{jpd} we notice that whenever $T$ increases (decreases) in $E$, the calculated correlation $r$ between $G$ map
and LDOS is negative (positive). This turns out to be a general rule observed also for larger energies and more subbands participating in the transport (see below).
For illustration, in the central panels of Figures \ref{jpo} and \ref{jpd} we plotted the sign of $r$ and the inverted sign of the derivative ${dT/dE}$.
Below, in subsection \ref{sec} we will provide an analytical argument for a perfect anticorrelation of $r$ and $dT/dE$ signs for strictly one-dimensional transport in
the perturbation theory \cite{pert1,pert2}. This finding in semi-classical WKB terms can be understood as follows: when $T$ grows with $E$,
a shorter electron wavelength is needed to tune the transfer probability to $T=1$.
The repulsive tip ($V_t>0$) makes the wavelengths  locally longer $\lambda(x,y)={2\pi \hbar}/{\sqrt{2m(E-V_t(x,y;x_t,y_t))}}$
which for $dT/dE>0$ lowers the value of $T$.

\begin{figure}[htbp]
\begin{centering}\begin{tabular}{lcr}
a)& & b) \\ & \includegraphics[bb=0 0 1300 1300, scale=0.20]{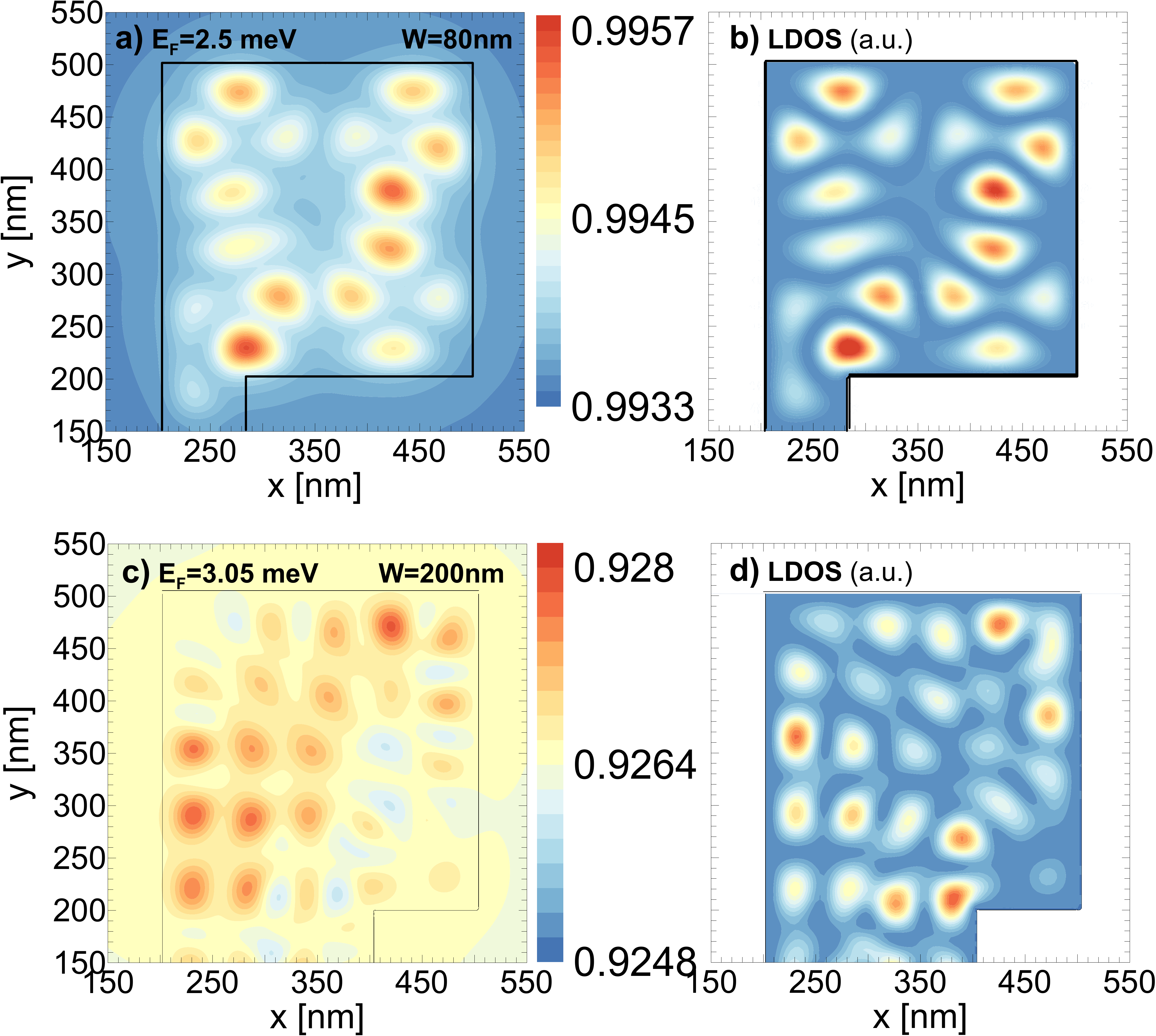}& \\
 c)& & d)
\end{tabular}
\end{centering}
\caption{Conductance maps (a,c) in units of $\frac{2e^2}{h}$ and the local density of states (b,d) for $W=50$ nm and $E=2.5$ meV (a,b)
and for $W=200$ and $E=3.05$ meV. Both cases correspond to $P=1$ with 1 and 4 channels of the dot to channel connection open.}\label{comp}
\end{figure}

Representative samples of $G$ maps and LDOS were plotted in Fig. \ref{comp}(a,b) for $W=80$ nm
and in Fig. \ref{comp}(c,d) for $W=200$ nm with energies 2.5 and 3.05 meV respectively [these work points are marked by dots in Fig. 3(d)].
Both energies  are taken off resonances and correspond to $dT(E)/dE<0$ [see Fig. \ref{jpo} and Fig. \ref{jpd}],
with positive $r$, equal to 0.93 and 0.42, respectively.
For the first choice [Fig. \ref{comp}(a,b)] we have a close correspondence of both maps as usual for the single mode of the connection.
For the latter some maxima of LDOS correspond to minima of $G$, hence the reduced value of $r$.

\subsection{Weak and strong coupling for $P>1$}

\begin{figure}[htbp]
\includegraphics[scale=0.5]{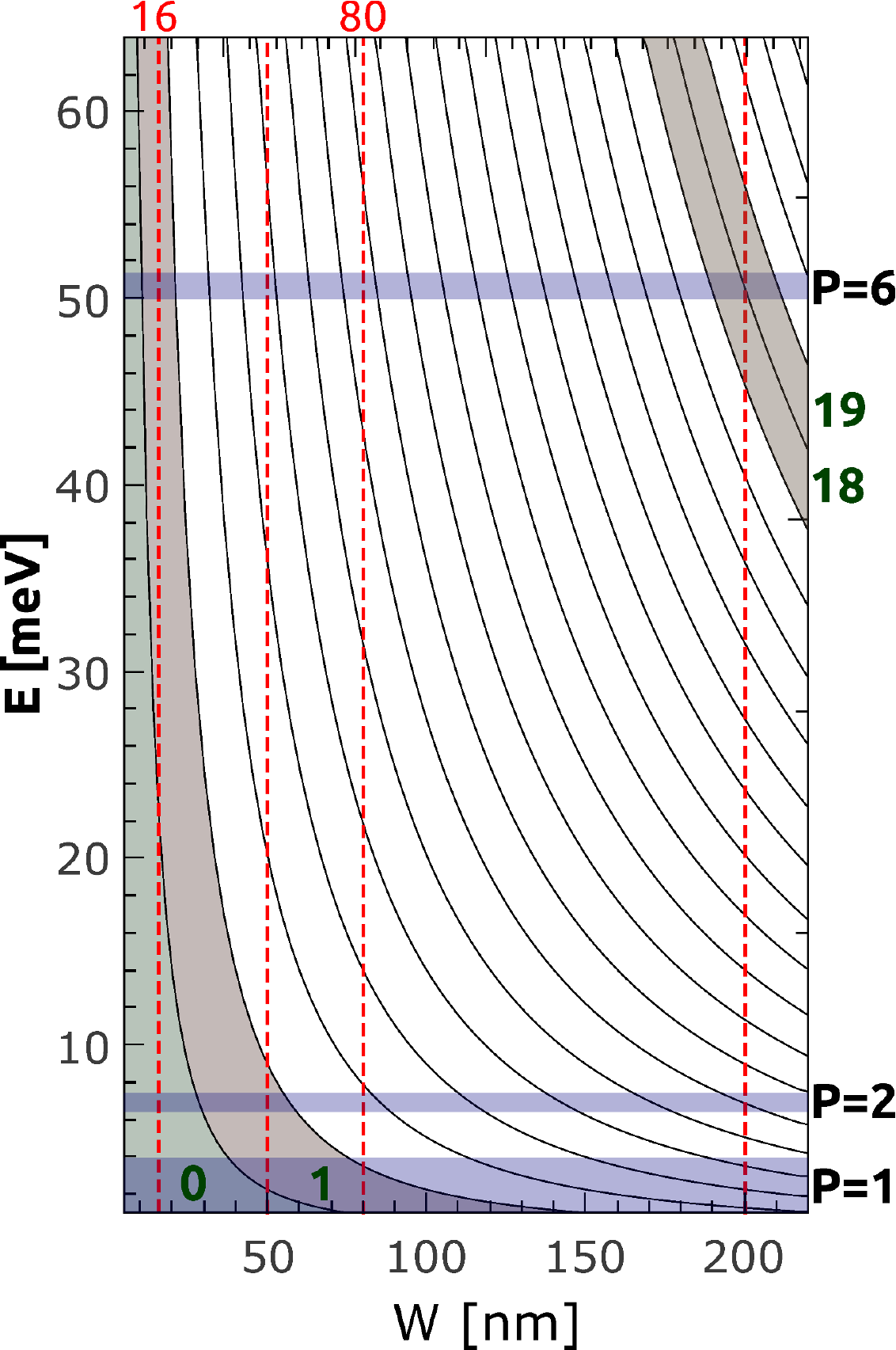}
\caption{Same as Fig. 3(d) only for wider $E$ range, including more subbands $P>1$.
The vertical lines show the values of $W$ considered in detail in this work. Horizontal
gray areas correspond to the energy ranges that we considered
in the $r(E)$ plots in this paper.  }\label{ms}
\end{figure}

In order to verify the general character of the above findings we performed
calculations for higher Fermi energy. The data will be discussed with respect to Fig. \ref{ms}
which shows the number of modes open in the connection for wider energy range than Fig. 3(d).

\begin{figure}[htbp]
\includegraphics[scale=0.5]{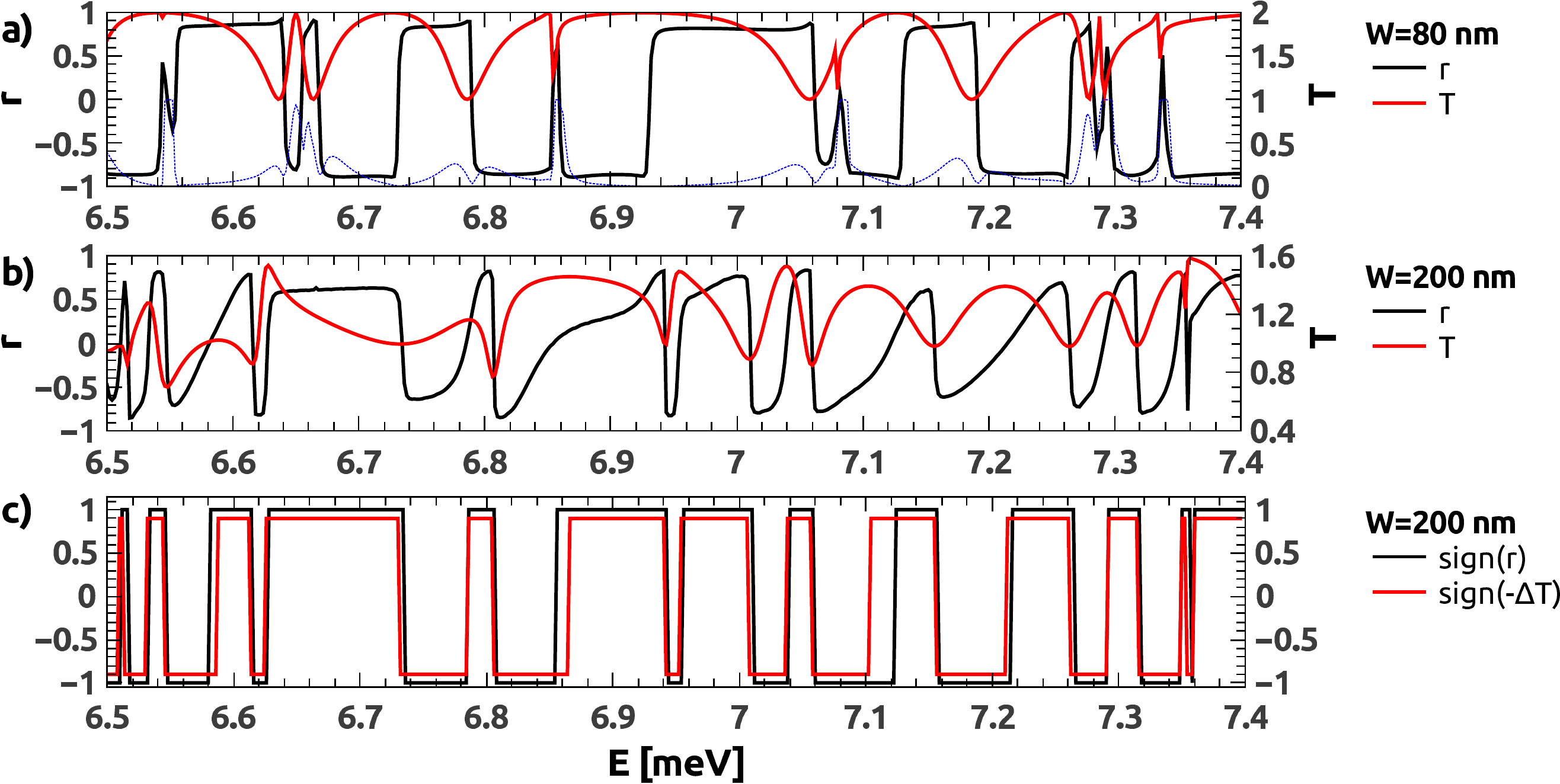}
\caption{ The energy range for $P=2$. The transfer probability and LDOS to $G$ correlation coefficient for $W=50$ nm (a) and $W=200$ nm (b).
 (c) The sign of the correlation $r$ and $-dT/dE$ (the latter multiplied by 0.9  for  clarity) for $W=200$ nm.
The blue line in (a) shows the contrast of the $G$ map (right axis).
}
 \label{2pww}
\end{figure}

Figure \ref{2pww} shows $T$ and $r$ for $P=2$ with
a single [$W=50$ nm, Fig. \ref{2pww}(a)] or multiple [$W=200$ nm, Fig. \ref{2pww}(b)] transparent modes in the connection
between the channel and the dot [see the gray belt for $P=2$ in Fig. \ref{ms}].
For the smaller value of $W$, $r$ has a binary dependence on $E$.
For the larger $W$ [see Fig. \ref{2pww}(b)] the variation of $r$ becomes smooth, and generally $|r|$ is far from 1, besides
the resonances.

\begin{figure}
\begin{centering}\begin{tabular}{c}
\includegraphics[scale=0.45]{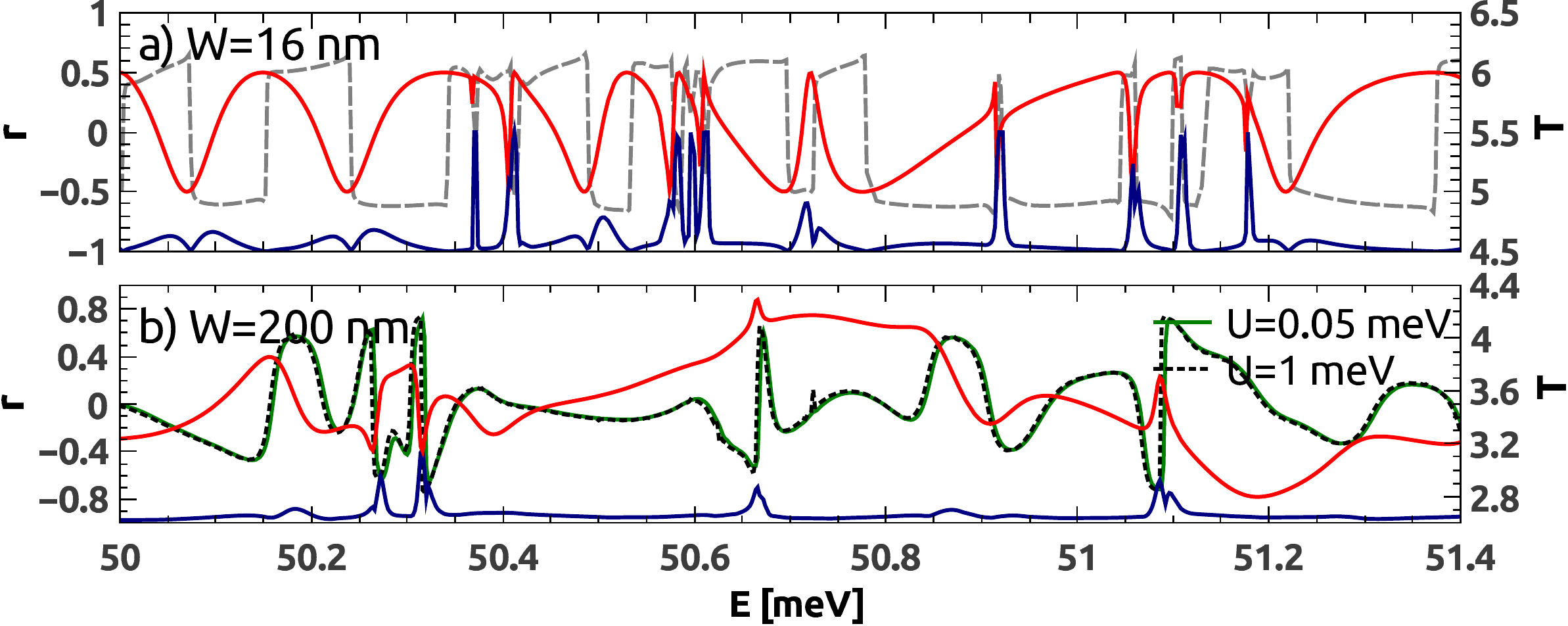}
\end{tabular}
\end{centering}
\caption{ Results for $P=6$ subbands at the Fermi level for $W=16$ nm (a) and $W=200$ nm (b).
For both (a) and (b) the red lines show  the transfer probability (right axis), the blue lines: the contrast,
here calculated as $T_{max}-T_{min}-1$ (left axis), and the correlation $r$ (black and grey curves, left axis). In (a) the contrast and $r$ were calculated for $U=1$ meV.
In (b) we show $r$ calculated for $U=0.05$ meV and $U=1$ meV (solid and dashed lines, respectively).
}
\label{p6}
\end{figure}

Results for six subbands at the Fermi level are given in Fig. \ref{p6}.
The binary dependence of $r$ is still obtained in spite of large $E$ but only for a very narrow opening with $W=16$ nm  -- see Fig. \ref{p6}(a).
This case corresponds to a single transparent mode of the connection -- cf. the gray belt with $P=6$ in Fig. \ref{ms}.
Similarly as seen for $P=2$ and $W=50$ nm in Fig. \ref{2pww}(a)
in the conditions of weak coupling, and the binary $r$ behavior we find $T(E)\in[P-1,P]$.
For $W=200$ nm at the resonances we obtain peaks of $|r|$ [Fig. \ref{p6}(b)]. Outside
the resonances $|r|$ remains small.

The contrast of the $G$ map calculated for $U=0.05$ meV is low outside the resonances [see Figures \ref{jpo} and \ref{jpd}].
However, the contrast can be increased by application of a larger potential with a limited influence
on the value of $r$. Figure \ref{p6}(b) compares the correlation coefficient calculated for $U=0.05$ meV (solid line) and for $U=1$ meV (dashed line).

\subsection{One dimensional case}
\label{sec}
\begin{figure}[htbp]
\begin{tabular}{cc}
\includegraphics[scale=0.6]{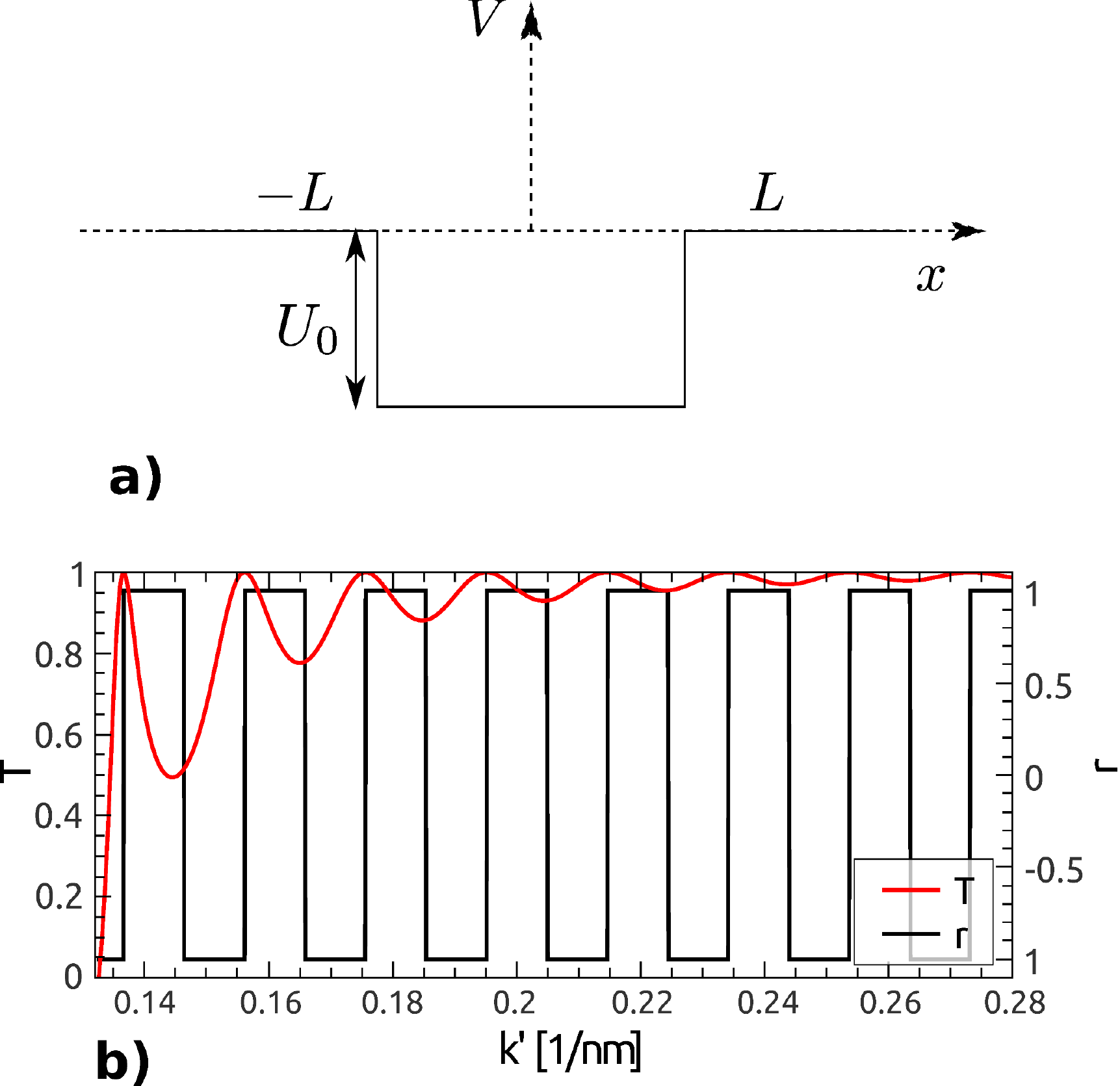}
\end{tabular}
\caption{(a) Quantum well of length $2L=200$ nm and depth $U_0=-10$ meV considered for the one-dimensional transport problem.
(b) Transfer probability (red line) given by Eq. (\ref{studnia}) and correlation of the LDOS with the $T$ maps (black curve).}
\label{qw}
\end{figure}

In the numerical results presented above we found that the sign of $r$ is anticorrelated to the sign of the derivative $dT/dE$.
It is possible to explain this finding analytically for a one-dimensional system.
Let us consider the quantum well of Fig. \ref{qw}(a). For the electron incident from the left, the scattering wave function
has the form $\phi_l=\exp(ikx)+c\exp(-ikx)$ for $x<-L$,
$$\phi_l = \frac{t}{2} e^{ikL}e^{-ik'L}(1+\frac{k}{k'})e^{ik'x}\left(1+e^{-2ik'x}e^{2ik'L}\frac{k'-k}{k'+k}\right)$$
for $x\in[-L,L]$
and $\phi_l=t\exp(ikx)$ for $x>L$, where $k=\sqrt{2mE}/\hbar$ and $k'=\sqrt{2m(E+U_0)}/\hbar$. For the symmetric quantum well a similar form of wave function is obtained for the electron incident from the right,
in particular inside the quantum well one has

$$\phi_r= \frac{t}{2} e^{ikL}e^{-ik'L}(1+\frac{k}{k'})e^{-ik'x}\left(1+e^{2ik'x}e^{2ik'L}\frac{k'-k}{k'+k}\right).$$

The local density of states within the quantum well is of the form \begin{equation} \mathrm{LDOS}=|\phi_l|^2+|\phi_r|^2= a'+b'\cos(2k'L)\cos(2k'x),\end{equation}
where $a'$ and $b'$ are variables which do not depend on the spatial coordinate $x$.

Let us now assume that the tip potential is of the delta form $V_t=U\delta(x-x_t)$.
For small $U$ the conductance change induced by the tip can be evaluated in the first order of the perturbation analysis of Ref. \cite{pert1,pert2}
\begin{equation} G(x_t)=G_0-4\pi \Im \left\{c^*t V_{21}(x_t)\right\},\end{equation}
with $V_{21}(x_t)=U \phi_r(x_t)^* \phi_l(x_t)$. The product  $$c^*t=i\frac{4kk'(k'-k)(k'+k)\sin(2k'L)}{(k'^{2}-k^{2})^{2}\cos(4k'L)-(k'^{2}+k^{2})^{2}-4k'^{2}k^{2}}$$ is purely imaginary.
Therefore we need only the real part of the matrix element $V_{21}$ which reads $\Re\left\{V_{21}(x_t\right\} \propto 2\beta\cos(2k'L)+(\beta^2+1)\cos(2k'x_t)$, where $\beta = \frac{k'-k}{k'+k}$.
In consequence the conductance as a function of the tip position $x_t$ has a form given by
\begin{equation}G(x_t)=a+b\sin(2k'L)\cos(2k'x_t),\end{equation}
where $a$ and $b$ do not depend on spatial coordinate $x$. Thus the differences between the maps of $G(x_t)$ and LDOS$(x_t)$ disappear when the maps are normalized for calculation
of the correlation $r$ [see Eq. (8)]. For a perturbative delta-like tip one finds exactly $|r|=1$ for any $E$ [see Fig. \ref{qw}(b)].
This is in a perfect agreement with the conclusion reached in Ref. \cite{pala2} for the discussion
of LDOS -- $G$ correlation in one dimension.

Now, lets us consider the sign of the correlation $r$. The transfer probability is given by
\begin{equation}
T=\frac{8k'^2k^2}{(k'^2+k^2)^2+4k'^2k^2-(k'^2-k^2)^2\cos(4k'L)},
\label{studnia}
\end{equation}
and reaches 1 when $k'=\frac{\pi n}{2L}$, for integer $n$. For these values of $k'$, for which $T=1$ the product $c^*t$
changes sign. Hence, at the peaks of $T$ the correlation $r$ changes sign.
The results for $T$ and $r$ are given in Fig. \ref{qw}(b) with $r=-1$ (+1) at the growing (decreasing) slope of $T$.

\section{Discussion}

The perfect correlation of LDOS with the $G$ map
is due to the inversional symmetry of the considered scattering potential. 
The 2D system studied in this work is not inversionally invariant, yet a good correspondence
is found for any $P$ provided that the channel to quantum dot connection is narrow.
Let us explain why the LDOS is resolved by $G$ maps for any $E$ when the dot is weakly  coupled to the channel.
For this purpose it is convenient to return to the single-subband ($P=1$) result of Fig. 3.

\begin{figure*}[htbp]
\includegraphics[bb =0 0  1480 580, scale=0.3]{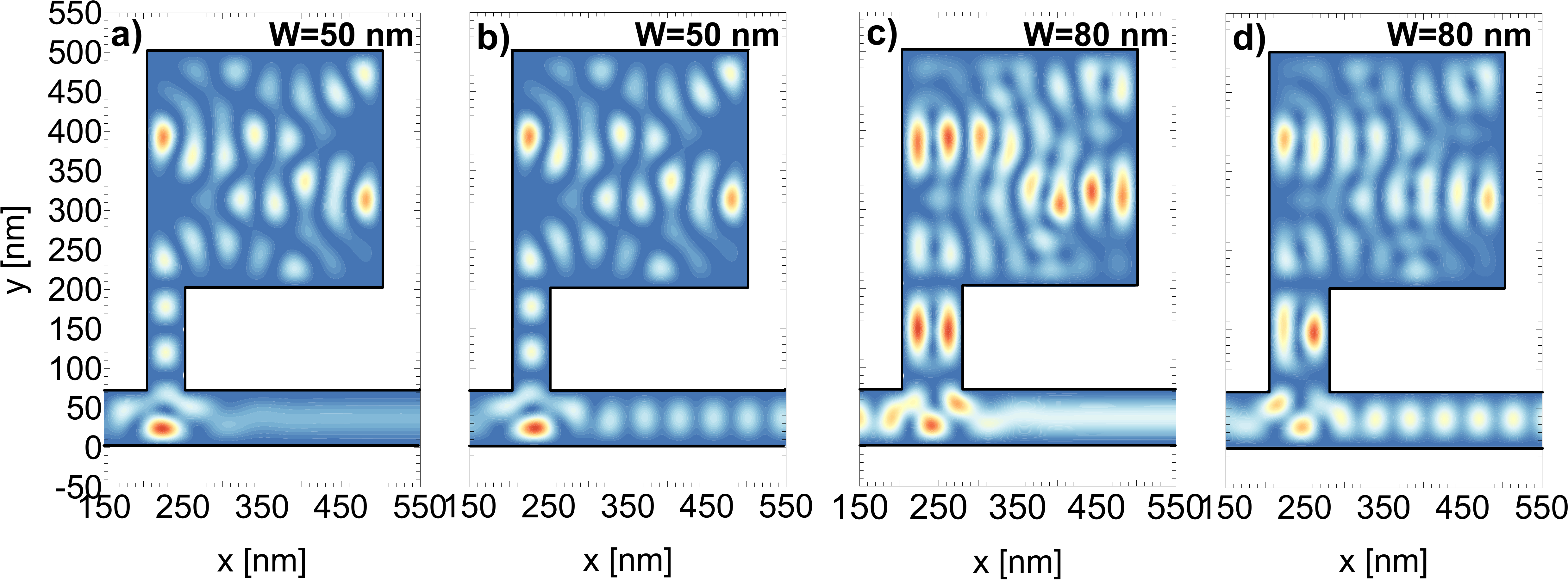}
\caption{Single subband transport $P=1$, for $E=4.1$ meV.
The probability density for the electron incident from the left (a,c)
and right (b,d) for $W=50$ nm (a,b) and $W=80$ nm (c,d).}\label{4w}
\end{figure*}

Figure \ref{4w}(a,c)
shows the scattering probability density
for the electron incident from the left [Fig. \ref{4w}] and right  [Fig. \ref{4w}(b,d)] for $E=4.1$ meV.
We can see that for $W=50$ nm -- for a single channel within the connection [cf. Fig. 3(d)] --
the densities inside the quantum dot are identical for both transport directions [Fig. \ref{4w}(a,b)].
For $W=50$ nm there is only a single open transport channel inside the connection. Thus the electron wave function on
its way to the quantum dot is bound to loose the information on the incidence direction.
This is no longer the case for $W=80$ nm [Fig. \ref{4w}(c,d)] for which two modes
of the connection participate in the transport [see Fig. 3(d) for $E=4.1$ meV] the values of $r$ loose their binary dependence on $E$.

For a single open mode of the transport across the connection -- not only the probability
density is identical for both incidences but also the wave function inside the quantum dot
becomes the same $\phi_r=\phi_l$ (with precision to a phase). Thus the matrix element of the perturbation theory
$V_{21}$  [see Eq. (10)] becomes proportional to the probability density and  it perfectly
matches the LDOS  which explains the central finding of the present work.

For $P>1$ the conditions for the binary values of $r(E)$  can be recognized
by the $T(E)$ dependence.
We have $P$ subbands carrying the current
 in the main channel of the system [Fig. 1]. In the discussed regime the connection to the quantum
dot can only transmit the current in a single mode. Hence,
the electron transfer probability summed over the incident subbands [Eq. (2)] should vary then between $(P-1)$ and $P$.
This is exactly what we found for binary $r(E)$ dependence with $P=2$ [see Fig. 8(a)]
and $P=6$ [Fig. 10(b)] subbands at the Fermi level.

\begin{figure}[htbp]
\includegraphics[bb =0 0  610 200, scale=0.6]{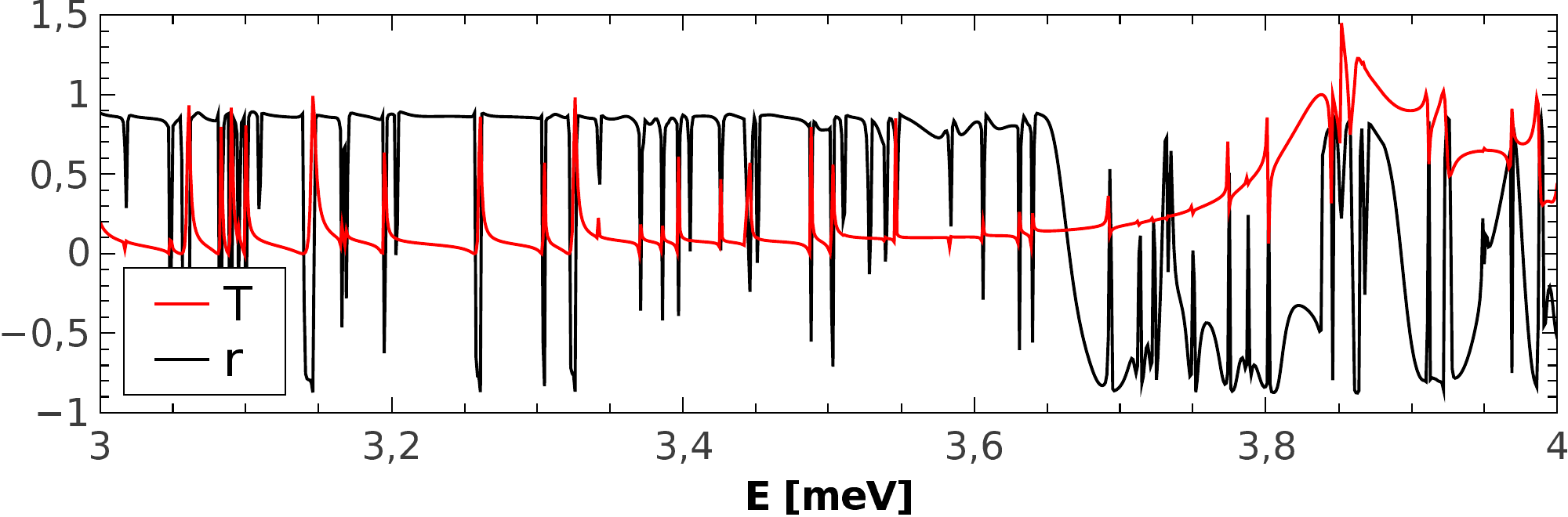}
\caption{Same as top panel of Fig. 4 but for the quantum dot of dimensions increased to 600 nm $\times$ 600 nm. The transfer probability $T$ and LDOS to $G$ correlation
factor $r$ for the energy range near the transition from a single to two conducting subbands of the connection $\simeq 3.7$ meV. The connection width is $W=80$ nm.}\label{ost}
\end{figure}

Let us comment on the experimental feasibility of the LDOS mapping for the weak coupling regime discussed in the present paper.
The open quantum dots connected to electron reservoirs are routinely produced by gating \cite{bird}, etching \cite{g6},
of surface oxidation \cite{fuhrer} techniques.  The open dots \cite{g6,bird,fuhrer} have usually size between 200 nm and 1 $\mu$m, i.e. of the
order of the dimension discussed in the present paper.
The dot needs to be connected to the reservoirs by a narrow channel. For low Fermi energy a channel of 80 nm is narrow enough.
For higher Fermi energies we discussed channels of width as small as 16 nm. 
Such narrow channels can be defined with the surface oxidation technique that uses the AFM tip to deposit the oxide
on specific locations of the sample. Formation of quantum wires of width as small as 4 nm by this technique was reported \cite{martinez}.
For good resolution of the $G$ maps the ratio of the size of the dot to the width of the tip should be the largest.
The correlation of the $G$ maps to the LDOS depends only on the width of the connection and stays high when
the quantum dot size is increased. Figure \ref{ost} shows the LDOS to $G$ map correlation factor for the quantum
dot size increased to  600 nm $\times$ 600 nm and $W=80$ nm. For $E$ below $\simeq 3.7$ meV the connection has a single conducting
channel and $r$ exhibits the binary dependence on the energy as for the case of the smaller dot (see Fig. 4).
On the other hand, the LDOS mapping can only be realized provided that the quantum dot size is smaller than
the coherence length. The direct measurements \cite{dmcl} of the phase coherence length in the GaAs two-dimensional electron gas.
indicate that the phase coherence length depends on the temperature $T$ as a $T^{-1/3}$ function,  and at $T=100$ mK the length is of the order of 1 $\mu$m.
The present results show that the $G$ map to LDOS correlation factor is a strongly varying function of the energy.
Besides the coherence length the temperature affects also the spread of the Fermi level. For $T=100$ mK, the spread $k_B T$ is of
the order of 10 $\mu$eV. This value is of the order of the width of resonances  [see Fig. 4]. However, outside the resonances for the weak
coupling, i.e. in conditions where  LDOS can be mapped for any $E$, variation of the transfer probability, the correlation factor and finally the LDOS with the Fermi energy is much slower,
so the LDOS imaging by $G$ measurements should be within the reach of the experiment.

\section{Summary and Conclusions}

We have studied the conductance response of an open quantum dot
to a short range perturbation that scans the surface of the system
in the context of mapping the local density of states.
The study covered up to $P=6$ subbands at the Fermi level
 and was supported by the analysis
of the density of states localized in the quantum dot on the energy scale. The latter turns out to
determine the contrast of the SGM conductance maps and the sign of the correlation coefficient $r$ between
the conductance maps and LDOS: the sign of $r$ is always opposite to the sign of derivative $dT/dE$.

We have found that for weak coupling of the quantum dot to the channel,
  the correlation coefficient of LDOS to $G$ maps $r$ takes binary values, which are close to either -1 or 1 for any $E$.
For strongly coupled quantum dots (large values of $W$) -- the dependence of $r(E)$ is no longer binary, but $r(E)$ tends to $\pm 1$
at the resonances -- for energies corresponding to the dot localized quasi-bound states.
We conclude that for weak coupling of the dot to the channel the  tip does not seem to interfere with the scattering
wave functions. In the conditions of weak coupling, the tip --  when localized near the maximum of LDOS -- tunes the scattering conditions to or away of an extremal value of $T$. We have explained that these special conditions -- when LDOS can be extracted from $G$ maps for an arbitrary Fermi energy -- appear when the connection of the channel to the quantum dot
allows only a single transport mode to pass from the channel to the dot.
Then, the wave functions inside the quantum dot for both incidence directions are the same up to a phase factor.
This implies that LDOS and conductance maps become identical
for a delta-like tip potential. We have indicated that an experimental signature of these conditions is the conductance changing
between $P-1$ and $P$ in units of $\frac{2e^2}{h}$ as the Fermi energy is changed.

\section*{Acknowledgements}
This work was supported by National Science Centre
according to decision DEC-2012/05/B/ST3/03290,
by the PL-Grid Infrastructure and by the Ministry of Science
and Higher Education within statutory tasks of the Faculty. Calculations were performed
in ACK -- CYFRONET -- AGH on the RackServer Zeus.


\begin{thebibliography}{00}
\bibitem{review1} Sellier H, Hackens B, Pala M G, Martins F, Baltazar
S, Wallart X, Desplanque L, Bayot V and Huant S, 2011, Sem. Sci. Tech.
\textbf{26}, 064008.

\bibitem{review2} Ferry D K, Burke A M, Akis R, Brunner R, Day T
E, Meisels R, Kuchar F, Bird J P, and Bennett B R, 2011, Sem. Sci.
Tech. \textbf{26}, 043001.

\bibitem{qpc0} Crook R, Smith C G, Simmons M Y, and Ritchie D A,
2000, Phys. Rev. B \textbf{62}, 5174.

\bibitem{qpc1} Topinka M A, LeRoy B J, Shaw S E J, Heller E J, Westervelt
R M, Maranowski K D, and Gossard A C, 2000, Science \textbf{289},
2323.

\bibitem{qpc2} Topinka M A, LeRoy B J, Westervelt R M, Shaw S E J,
Fleischmann R, Heller E J, Maranowski K D, Gossard A C, 2001, Nature,
\textbf{410}, 183.

\bibitem{qpc3} Aidala K E, Parott R E, Kramer T, Heller E J, Westervelt
R M, Hanson M P, and Gossard A C, 2007, Nat. Phys. \textbf{3}, 464.

\bibitem{qpc4} Jura M P, Topinka M A, Urban L, Yazdani A, Shtrikman
H, Pfeiffer L N, West K W, and Goldhaber-Gordon D, 2007, Nat. Phys.
\textbf{3}, 841.

\bibitem{qpc5} Aoki N, Da C R, Akis C R, Ferry D K, and Ochiai Y,
2005, Appl. Phys. Lett. \textbf{87}, 223501.

\bibitem{qpc6} Pioda A, Kicin S, Brunner D, Ihn T, Sigrist M, Ensslin
K, Reinwald M, and Wegscheider W, 2007, Phys. Rev. B \textbf{75},
045433.

\bibitem{qpc7} Schnez S, Rossler C, Ihn T, Ensslin K, Reichl C, and
Wegscheider W, 2011, Phys. Rev. B \textbf{84}, 195322.

\bibitem{qpc8} Kozikov A A, Weinmann D, R\"ossler C, Ihn T, Ensslin
K, Reichl C, and Wegscheider W, 2009, New J. Phys. \textbf{15}, 083005.

\bibitem{qh1} Hackens B, Martins F, Faniel S, Dutu C A, Sellier
H, Huant S, Pala M, Desplanque L, Wallart, X and Bayot V, 2010,
Nat. Commun. \textbf{1}, 39.

\bibitem{qh2} Martins F, Faniel S, Rosenow B, Pala M G, Sellier
H, Huant S, Desplanque L, Wallart X, Bayot V, Hackens B, 2013,
New J. Phys. \textbf{15}, 013049.

\bibitem{edge} Rychen J, Van\v{c}ura T, Ensslin K, Wegscheider
W, Bichler M, 2005, Physica E \textbf{13}, 671(2002); Aoki N, da
Cunha C R, Akis R, Ferry D K, Ochiai Y, 2005, Phys. Rev. B \textbf{72},
155327.

\bibitem{pala1} Martins F, Hackens B, Pala M G, Ouisse T, Sellier
H, Wallart X, Bollaert S, Cappy A, Chevrier J, Bayot V, and
Huant S, 2007, Phys. Rev. Lett. \textbf{99}, 136807.

\bibitem{pala2} Pala M G, Hackens B, Martins F, Sellier H, Bayot
V, Huant S, and Ouisse T, 2008, Phys. Rev. B \textbf{77}, 125310.

\bibitem{pala3} Pala M G, Baltazar S, Martins F, Hackens B,
Sellier H, Ouisse T, Bayot V, and Huant S, 2009, Nanotechnology
\textbf{20}, 264021.

\bibitem{szafran} Szafran B 2011 Phys. Rev. B \textbf{84}, 075336.

\bibitem{chwiej} Chwiej T and Szafran B, 2013, Phys. Rev. B \textbf{87},
085302.

\bibitem{petrovic} Petrovic M D, Peeters F M, Chaves A, and Farias
G A, 2013, J. Phys.:Condens. Matter \textbf{25}, 495301.

\bibitem{qdc1} Fallahi P, Bleszynski A C, Westervelt R M, Huang
J, Walls J D, Heller E J, Hanson M, and Gossard A C, 2005, Nano
Lett. \textbf{5}, 223; Zhang L Mand Fogler M M, 2006, Nano Lett.
\textbf{6}, 2206; Gildemeister A E, Ihn T, Sigrist M, and Ensslin
K, 2007, Phys. Rev. B \textbf{75}, 195338; Bleszynski-Jayich A C,
Fr\"oberg L E, Bj\"ork M T, Trodahl H J, Samuelson L, and Westervelt
R W, 2008, Phys. Rev. B \textbf{77}, 245327; Qian J, Halperin B I,
and Heller E J, 2010, Phys. Rev. B \textbf{81}, 125323; Boyd E E
and Westervelt R M, 2011, Phys. Rev. B \textbf{84}, 205308; Boyd
E E, Storm K, Samuelson L, and Westervelt R M, 2011, Nanotechnology
\textbf{22}, 185201. Huefner M, Kueng B, Schnez S, Ensslin K,
Ihn T, Reinwald M and Wegscheider W, 2011, Phys. Rev. B \textbf{83},
235326. Mantelli D, Cavaliere F, and Sassetti M, 2012, J. Phys.
Condens. Matter \textbf{24}, 43202. Ziani N T, Caveliere F, and Sassetti
M, 2012, Phys. Rev. B \textbf{86}, 125451.

\bibitem{shulz1} Mendoza M and Schulz P A, 2003, Phys. Rev. B \textbf{68},
205302.

\bibitem{shulz2} Mendoza M and Schulz P A, 2005, Phys. Rev. B \textbf{71},
245303.

\bibitem{qda1} Ferry D K, Akis R, and Bird J P, 2004, Phys. Rev.
Lett. \textbf{93}, 026803.

\bibitem{g6} Burke A M, Akis R, Day T E, Speyer G, Ferry D K,
and Bennett B R, 2010, Phys. Rev. Lett. \textbf{104}, 176801.

\bibitem{datta} Datta S, Electronic Transport in Mesoscopic Systems,
1995, Cambridge Univ. Press, Cambridge.


\bibitem{kramer} Kramer S, 2013, Phys. Rev. B \textbf{88}, 125308.

\bibitem{bruno} Metalidis G and Bruno P, 2005, Phys. Rev. B \textbf{72},
235304.

\bibitem{cresti} Cresti A, 2006, J. Appl. Phys. \textbf{100}, 053711.

\bibitem{pert1} Jalabert R A, Szewc W, Tomsovic S, and Weinmann
D, 2010, Phys. Rev. Lett. \textbf{105}, 166802.

\bibitem{pert2} Gorini C, Jalabert R A, Szewc W, Tomsovic S,
and Weinmann D, 2013, Phys. Rev. B \textbf{88}, 035406.

\bibitem{abbout} Abbout A, Lemarie G, and Pichard J L, 2011,
Phys. Rev. Lett. \textbf{106}, 156810.

\bibitem{jura} Jura M P, Topinka M A, Grobis M, Pfeiffer L N, West
K W, and Goldhaber-Gordon D, 2009, Phys. Rev. B \textbf{80}, 041303(R).

\bibitem{kozikov} Kozikov A A, R\"ossler C, Ihn T, Ensslin K, Reich C, 2013, New J. Phys. \textbf{15}, 013056.

\bibitem{kolasinski} Kolasi\'{n}ski K and Szafran B, 2013, Phys.
Rev. B \textbf{88}, 165306.

\bibitem{shulz3} Mendoza M, Schulz P A, Vallejos R O,and Lewenkopf
C H, 2008, Phys. Rev. B \textbf{83}, 155307.

\bibitem{morfonios} Morfonios C, Buchholz D, and Schmelcher P,
2011, Phys. Rev. B \textbf{83}, 205316.

\bibitem{clerk} Clerk A A, Waintal X, and Brouwer P W, 2001,
Phys. Rev. Lett. \textbf{86}, 4636.

\bibitem{Barn} Barnthaler A, Rotter S, Libisch F, Burgd\"orfer J,
Gehler S, Kuhl U, and Stockmann H -J, 2010, Phys. Rev. Lett. \textbf{105},
056801.

\bibitem{Miro} Miroshnichenko A E, Flach S, and Kivshar Y S,
2010, Rev. Mod. Phys. \textbf{82}, 2257.

\bibitem{governale} Governale M, and Ungarelli C, 1998, Phys. Rev.
B \textbf{58}, 7816.

\bibitem{mandel} Mandelshtam V A, Ravuri T R, Taylor H S, 1993,
Phys. Rev. Lett. \textbf{70}, 1932.


\bibitem{akis} Akis R, Vasilopoulos P, and Debray P, 1997, Phys.
Rev. B \textbf{56}, 9594.

\bibitem{nockel} N\"ockel J U, 1992, Phys. Rev. B \textbf{46}, 15348.

\bibitem{bird} Bird J P, Akis R, Ferry D K, Vasieleska D, Cooper J, Aoyagi Y, Sugano T, 1999,
Phys. Rev. Lett. {\bf 82}, 4691.

\bibitem{fuhrer} Fuhrer A, L\"uscher S, Ihn T, Heinzel T, Ensslin K, Wegscheider W, and Bichler M, 2001
Nature {\bf 413}, 822.

\bibitem{martinez} Martines J, Martinez R V, and Garcia R, 2008, Nano Lett. {\bf 8} 3636.

\bibitem{dmcl} Ferrier M, Angers L, Rowe A C H, Gue\'eron, Bouchiat H, Texier C, Montambaux G, Mailly D, 2004
Phys. Rev. Lett. {\bf 93}, 246804.

\end{thebibliography}
\end{document}